\documentclass[12pt]{article}
\usepackage{epsfig,graphicx}
\usepackage{fpcp04}
\usepackage{amsmath}
\usepackage{amssymb}
\usepackage{array}
\usepackage{ulem}

\def\beq{\begin{equation}}
\def\eeq#1{\label{#1}\end{equation}}
\def\eeqn{\end{equation}}

\def\beqa{\begin{eqnarray}}
\def\eeqa#1{\label{#1}\end{eqnarray}}
\def\eeqan{\end{eqnarray}}

\let\bar=\overbar

\def\vev#1{\langle #1 \rangle}

\def\Dslash{\not{\hbox{\kern-4pt $D$}}}
\def\dslash{\not{\hbox{\kern-2pt $\del$}}}

\def\msb{{\bar{\ssstyle M \kern -1pt S}}}

\def\vckm{V_{\rm CKM}}

\def\BB0bar{B^0 {\overline B}^0}
\def\BB0dbar{B_d^0 {\overline B}_d^0}
\def\BB0sbar{B_s^0 {\overline B}_s^0}

\RequirePackage{xspace}

\usepackage{relsize}
\def\babar{\mbox{\slshape B\kern-0.1em{\smaller A}\kern-0.1em
    B\kern-0.1em{\smaller A\kern-0.2em R}}}

\def\d     {\ensuremath{d}\xspace}

\def\Kbar  {\kern 0.2em\overline{\kern -0.2em K}{}\xspace}

\def\Kz    {\ensuremath{K^0}\xspace}
\def\Kzb   {\ensuremath{\Kbar^0}\xspace}
\def\KzKzb {\ensuremath{\Kz \kern -0.16em \Kzb}\xspace}
\def\Kp    {\ensuremath{K^+}\xspace}
\def\Km    {\ensuremath{K^-}\xspace}

\def\KpKm  {\ensuremath{\Kp \kern -0.16em \Km}\xspace}

\def\Dbar    {\kern 0.2em\overline{\kern -0.2em D}{}\xspace}

\def\Dz      {\ensuremath{D^0}\xspace}
\def\Dzb     {\ensuremath{\Dbar^0}\xspace}
\def\DzDzb   {\ensuremath{\Dz {\kern -0.16em \Dzb}}\xspace}
\def\Dp      {\ensuremath{D^+}\xspace}
\def\Dm      {\ensuremath{D^-}\xspace}

\def\DpDm    {\ensuremath{\Dp {\kern -0.16em \Dm}}\xspace}

\def\Bbar    {\kern 0.18em\overline{\kern -0.18em B}{}\xspace}

\def\BB      {\ensuremath{B\Bbar}\xspace} 
\def\Bz      {\ensuremath{B^0}\xspace}
\def\Bzb     {\ensuremath{\Bbar^0}\xspace}
\def\BzBzb   {\ensuremath{\Bz {\kern -0.16em \Bzb}}\xspace}
\def\Bu      {\ensuremath{B^+}\xspace}
\def\Bub     {\ensuremath{B^-}\xspace}

\def\BpBm    {\ensuremath{\Bu {\kern -0.16em \Bub}}\xspace}

\mathchardef\Upsilon="7107
\def\Y#1S{\ensuremath{\Upsilon{(#1S)}}\xspace}

\mathchardef\Deltares="7101
\mathchardef\Xi="7104
\mathchardef\Lambda="7103
\mathchardef\Sigma="7106
\mathchardef\Omega="710A

\def\Deltabar{\kern 0.25em\overline{\kern -0.25em \Deltares}{}\xspace}
\def\Lbar{\kern 0.2em\overline{\kern -0.2em\Lambda\kern 0.05em}\kern-0.05em{}\xspace}
\def\Sigbar{\kern 0.2em\overline{\kern -0.2em \Sigma}{}\xspace}
\def\Xibar{\kern 0.2em\overline{\kern -0.2em \Xi}{}\xspace}
\def\Obar{\kern 0.2em\overline{\kern -0.2em \Omega}{}\xspace}
\def\Nbar{\kern 0.2em\overline{\kern -0.2em N}{}\xspace}
\def\Xb{\kern 0.2em\overline{\kern -0.2em X}{}\xspace}

\newcommand{\tev}{\ensuremath{\mathrm{\,Te\kern -0.1em V}}\xspace}
\newcommand{\gev}{\ensuremath{\mathrm{\,Ge\kern -0.1em V}}\xspace}
\newcommand{\mev}{\ensuremath{\mathrm{\,Me\kern -0.1em V}}\xspace}
\newcommand{\kev}{\ensuremath{\mathrm{\,ke\kern -0.1em V}}\xspace}
\newcommand{\ev}{\ensuremath{\mathrm{\,e\kern -0.1em V}}\xspace}
\newcommand{\gevc}{\ensuremath{{\mathrm{\,Ge\kern -0.1em V\!/}c}}\xspace}
\newcommand{\mevc}{\ensuremath{{\mathrm{\,Me\kern -0.1em V\!/}c}}\xspace}
\newcommand{\gevcc}{\ensuremath{{\mathrm{\,Ge\kern -0.1em V\!/}c^2}}\xspace}
\newcommand{\mevcc}{\ensuremath{{\mathrm{\,Me\kern -0.1em V\!/}c^2}}\xspace}

\def\mus  {\ensuremath{\rm \,\mus}\xspace}

\def\mus        {\ensuremath{\,\mu{\rm s}}\xspace}    

\def\to                 {\ensuremath{\rightarrow}\xspace}

\def\pep2{PEP-II}

\def\gsim{{~\raise.15em\hbox{$>$}\kern-.85em
          \lower.35em\hbox{$\sim$}~}\xspace}
\def\lsim{{~\raise.15em\hbox{$<$}\kern-.85em
          \lower.35em\hbox{$\sim$}~}\xspace}

\def\Vtd  {\ensuremath{|V_{td}|}\xspace}

\def\Vub  {\ensuremath{|V_{ub}|}\xspace}
\def\Vcb  {\ensuremath{|V_{cb}|}\xspace}

\xspace

\def\jetset74   {\mbox{\tt Jetset \hspace{-0.5em}7.\hspace{-0.2em}4}\xspace}

\newcommand{\slideframe}[1]{{}}

\newcommand{\bei}{\begin{itemize}}
\newcommand{\eei}{\end{itemize}}

\def\OMIT#1{{}}

\newcommand{\LQCD}{{\Lambda_{\rm QCD}}}
\newcommand{\texthalf}{{\textstyle{\frac12}}} 
\newcommand\spur{\raise.15ex\hbox{/}\kern-.57em }

\newcommand{\CO}{\mathcal{O}}

\newcommand{\ecut}{{E_{\text{cut}}}}
\def\d{{\rm d}}
\def\FD{{\cal F}}
\def\FDs{\FD_*}

\begin{document}

\Title{CKM Sides: Theory}
\bigskip

%
\label{GrinsteinStart}

%
\author{Benjam\'\i{}n Grinstein}

%
\address{Department of Physics\\
University of California, San Diego\\
La Jolla, CA 92093-0319, USA\\
}

\makeauthor\abstracts{ We review the theory of the determination of
the CKM elements $\Vcb$ and $\Vub$.  Particular attention is paid to
the determination of $\Vcb$ through inclusive semileptonic $B$ decays
to charm using a moment analysis, since this
has shown most progress recently. A precise method for the determination
of $\Vub$ via exclusive decays is described.}

\section{Introduction}
The Wolfenstein parametrization of the CKM matrix,
\beq 
\vckm=\begin{pmatrix}
V_{ud}&V_{us}&V_{ub}\\
V_{cd}&V_{cs}&V_{cb}\\
V_{td}&V_{ts}&V_{tb}
\end{pmatrix}
\approx \begin{pmatrix}
1-\texthalf\lambda^2&\lambda&A\lambda^3(\bar\rho-i\bar\eta)\\
-\lambda(1+iA^2\lambda^4\bar\eta & 1-\texthalf\lambda^2& A\lambda^2\\
 A\lambda^3(1-\bar\rho-i\bar\eta)& -A\lambda^2(1+i\lambda^2\bar\eta)& 1
\end{pmatrix} + \mathcal{O}(\lambda^6)
\eeqn
shows explicitly that it is determined by four parameters. $K\ell3$
decays determine $\lambda$ to 1\% accuracy\cite{glazov}. The other
three parameters are the subject of this talk. As we will see,
$|V_{cb}|=A\lambda^2$ is now determined to nearly 1\% accuracy from
$B\to X_c\ell\nu_\ell$. The magnitude of the remaining complex
parameter, $\bar\rho-i\bar\eta=V_{ub}/(V_{us}V_{cb})$, is fixed from
$B\to X_u\ell\nu_\ell$, but is known far less accurately. 

Joining the points 0, 1 and $\bar\rho+i\bar\eta$ in the complex plane
yields a {\sl unitarity triangle.} The sides have length 1,
$\sqrt{\bar\rho^2+\bar\eta^2}=|V_{ub}|/|V_{us}V_{cb}|$ and
$\sqrt{(1-\bar\rho)^2+\bar\eta^2}=|V_{td}|/|V_{us}V_{cb}|$. To
determine the triangle one needs not just the magnitude but also the
phase of $\bar\rho-i\bar\eta$. Alternatively, the length of the third
side determines the triangle, up to a two-fold ambiguity.
This third side can be obtained from $b\to d\gamma$
decays (eg, $B\to\rho\gamma$) or from the $B^0-\overline {B^0}$ mixing
parameter $\Delta M_d$. In the standard model, both of these processes
are dominated by a virtual top quark exchange and depend directly on
$V^{\phantom{*}}_{td}V^*_{tb}$. The theory of these  processes belongs
in a different session in this conference\cite{rare1,rare2}.

The bulk of the talk is devoted to the determination of $|V_{cb}|$
from inclusive decays $B\to X_c\ell\nu_\ell$, since here is where we
have seen tremendous progress in the last year. Next we briefly
revisit $|V_{cb}|$ from exclusive decays, $B\to D^*\ell\nu_\ell $ and
$B\to D\ell\nu_\ell $. We then turn to $|V_{ub}|$ and mirroring the
discussion above we first present its determination from inclusive
decays, $B\to X_u\ell\nu_\ell$. This is followed by a review of the
extraction of $|V_{ub}|$ from exclusives $B\to \pi(\rho)
\ell\nu_\ell$. The limitted precission in the determination of $\Vub$
afforded by the standard methods calls for innovation in this area. We
describe a new method that uses a combination of radiative and
semileptonic $B$ and $D$ decays.

\section{$|V_{cb}|$ from Inclusive decays $B\to X_c\ell\nu_\ell$}
\label{sec:b->c}
In QCD the differential width for $B\to X_c\ell\nu_\ell$ is a function
of $|V_{cb}|$, the strong coupling $\alpha_s$ and the quark masses
$m_b$ and $m_c$. Only the dependence on $|V_{cb}|$ is trivial: 
\beq 
\frac{d\Gamma(B\to X_c\ell\nu_\ell)}{dx\,dy}={ |V_{cb}|^2} 
f(x,y;m_{b,c},\alpha_s),
\eeqn
where $x,y$ are kinematic variables. Even if we could compute the
function $f$ from first principles, we would still need as input
precise values of the three parameters, $m_{b}$, $m_c$ and $\alpha_s$. While
$\alpha_s$ is accurately known\cite{PDG}, the quark masses are
not. For a determination of $V_{cb}$ one can use the measurement of
the doubly differential decay rate to fit also the value of the
masses. 

Now, $V_{cb}$ enters only the overall normalization, while $m_{b}$ and $m_c$
determine the ``shape'' ({\it i.e.},\/ the functional form) of the decay
distribution. So one can isolate the determination of the masses by
fitting to shape parameters, that is, various moments of the
distribution. In all cases the moments are defined with a charged
lepton energy cut, an experimental necessity turned into a theoretical
tool. From the charged lepton integrals with an  energy cut
\beq
R_n(\ecut,M)\equiv \int_\ecut\!\!\!(E_\ell-M)^n
\frac{\d\Gamma}{\d E_\ell}\,\d E_\ell\,,
\eeqn
define moments
\beq
\vev{E^n_\ell}_\ecut\equiv \frac{R_n(\ecut,0)}{R_0(\ecut,0)}\,.
\eeqn
Central moments are also defined:
\beq
\vev{(E_\ell-\vev{E_\ell}_\ecut)^n}_\ecut
=\frac{R_n(\ecut,\vev{E_\ell}_\ecut)}{R_0(\ecut,0)}\,.
\eeqn
Also useful are moments of the hadron invariant mass, 
\(
m_X^2\equiv (p_B-p_\ell-p_\nu)^2=(p_B-q)^2
\), defined by
\beq
\vev{m_X^{2n}}_\ecut\equiv
\frac{\int_\ecut\!(m_X^2)^n\frac{\d\Gamma}{\d m^2_X}\,\d m^2_X}%
{\int_\ecut\!\frac{\d\Gamma}{\d m^2_X}\,\d m^2_X}\,.
\eeqn
Central moments $\vev{(m_X^2-{\overline m}^2_D)^n}$, about \(
{\overline m}^2_D\equiv\left[({m_D+3m_{D^*}})/4\right]^2 \) are also
frequently considered. Somewhat orthogonal information can be garnered
from moments of the $B\to X_s\gamma$ spectrum, which depend on $m_b$
and $\alpha_s$, and only very weakly on $m_c$. They are defined for
the spectrum of the photon energy in the $B$ rest frame, $E_\gamma$:
 \beq \vev{E^n_\gamma}\equiv
\frac{\int_\ecut\!(E_\gamma)^n
\frac{\d\Gamma}{\d E_\gamma}\,\d E_\gamma}
{\int_\ecut\!
\frac{\d\Gamma}{\d E_\gamma}\,\d E_\gamma} 
\eeqn
The variance, $\vev{E_\gamma^2}-\vev{E_\gamma}^2$, is often used, but 
higher moments are not, because they are very sensitive to the boost of
the $B$ meson and  to details of shape function.

\subsection{Heavy Quark/Mass Expansion}
At present we don't know how to perform the calculation described
above. Instead a systematic approximation to QCD is made by expanding
in inverse powers of the heavy mass, $m_b$, and performing an operator
product expansion\cite{chay}. The price we pay is that new
dimensionful parameters are introduced. At $n$-th order in the
expansion the new parameters are roughly of size $(\LQCD)^n$, giving
rise to an expansion in powers of $(\LQCD/m_b)$. These new parameters
account for our ignorance of non-perturbative dynamics in QCD.

There are two approaches to the expansion in the literature:
\begin{itemize}
\item {\uline\sl The HQET/OPE}, in which amplitudes are given in terms of
  mass independent heavy meson  states. If one chooses to treat the
  $D^{(*)}$-meson as heavy then this approach involves an expansion in
  $1/m_c$ (in addition to $1/m_b$). The advantage is that $m_c$ can be
  fixed in terms of $m_b$ using the accurately determined  $M_D-M_B$.
\item {\uline\sl The HME}, or Heavy Mass Expansion, in which heavy meson states
  are not expanded (in neither $m_c$ nor $m_b$). No $1/m_c$ expansion
  is necessary. However, $m_c$ is a parameter determined from the fit.
\end{itemize}

Three groups have performed a moment analysis. The groups of
M. Battaglia {\it et al}\cite{battaglia} and of Gambino and
Uraltsev\cite{gambino} use the HME, while 
C. Bauer  {\it et al}\cite{bauer} consider and compare both
expansions. In all cases the expansion is up to (and including) terms of
    $\mathcal{O}(\LQCD)^3$, and  known perturbative corrections  are
included.  More on this below.

\subsection{Parameter Counting}
It may come as a surprise that the number of parameters introduced by
the two approaches is the same. The naive guess that the HQET/OPE with
a $1/m_c$ expansion ought to introduce more parameters because it
expands in $1/m_c$ too is incorrect. The additional expansion gives
new information, for example, that to leading order the  $B$
and $D$ meson states form heavy flavor symmetry doublets.

Altogether, there are six parameters (in addition to $\alpha_s$) that
enter the moment analysis in either approach.  With a $1/m_c$
expansion $m_c$ is fixed using $M_B-M_D$, and the parameters
$\lambda_2\sim(\LQCD)^2$ and $\rho_2\sim(\LQCD)^3$ are fixed from
$M_{B^*}-M_B$ and $M_{D^*}-M_D$. One is left with $m_b$,
$\lambda_1\sim(\LQCD)^2$, and four parameters of order $(\LQCD)^3$,
namely $\rho_1$ and three linear combinations of four time ordered
products, $\mathcal{T}_{1-4}$. Now one has the issue of accuracy of
expanding in $1/m_c$. The OPE is an expansion in $1/m_b$. It is only
the computation of $m_c$ in terms of meson masses that requires a
$1/m_c$ expansion. Since this is done to $\mathcal{O}(\LQCD/m_c)^3$,
the fractional error in $m_c$ is $\mathcal{O}(\LQCD/m_c)^4$. But $m_c$
first enters the rate at $\mathcal{O}(m_c/m_b)^2$, so the error
introduced is $\mathcal{O}(\Lambda^4_{\text{QCD}}/m_c^2m_b^2)$. 

If the states are not expanded in $1/m$ (so no $1/m_c$ expansion is
performed) then no time ordered products appear. However, $m_c$,
$\lambda_2$ and $\rho_2$ cannot be solved in terms of physical meson
masses, and they come in as parameters. Comparing with the counting
above, the three combinations of time order products   are
replaced here by the three parameters $m_c$, $\lambda_2$ and
$\rho_2$. Strictly speaking the parameters should be distinguished
from the ones above and denoted by different symbols since in the HME
the non-perturbative parameters have implicit dependence on the heavy
quark mass. Bauer {\it et al} suggest that the fit to moments may be
better behaved when the $1/m_c$ expansion is performed, since more of
the parameters are of higher order in $\LQCD$.

In either approach one needs to properly define quark masses. A good
definition of the quark mass will give better convergence of the
perturbatively calculable (in $\alpha_s$) Wilson coefficients of the
OPE. Many definitions are available\cite{luke}. Threshold mass
definitions, like the 1S and PS, give better convergence than the
pole, $\overline{\text{MS}}$ or kinetic mass schemes. Battaglia {\it et al}
and Gambino and Uraltsev use the kinetic mass scheme, while Bauer et
al use all and compare. This latter group points out a problem with
the kinetic mass. Since this scheme requires a rather low hard cut-off
$\mu$, when there are two heavy quark masses, $m_b$ and $m_c$ one
introduces two distinct hard cut-offs, $\mu_b$ and $\mu_c$. While it
is customary to use $\mu_b=\mu_c$, there is nothing in principle which
requires one to use equal cut-offs for the two masses. Bauer {\it et al}
consider the effect of changing $\mu_{b,c}$ so that $\mu_b\neq\mu_c$,
and find rather large variations in the determination of
$|V_{cb}|$. The effect is due, presumably, to the next-order
corrections in $\alpha_s(\mu_{b,c})$, which are rather large given
that $\mu_{b,c}$ must be taken to be of the order of the fairly low
scale $\sim1$~GeV.

\subsection{Fit results}

The BaBar Collaboration has performed a fit to the moments calculated
by Gambino and Uraltsev. 
The electron energy moments are computed to
order $\beta_0\alpha_s^2$, $\alpha_s(\LQCD)^2$ and $(\LQCD)^3$, while
the hadronic mass moments are computed to order $\alpha_s$,
$\alpha_s(\LQCD)^2$ and $(\LQCD)^3$. To account for the missing order
$\beta_0\alpha_s^2$ effects, the analysis uses a different value of
$\alpha_s$ for the hadronic mass moments than for the lepton energy
moments. This educated guess is somewhat {\it ad hoc} and introduces
an unknown uncertainty.  The result of the fit is presented in this
conference elsewhere\cite{elisabetta}. The fit includes non-integral
moments of $m^2_X$, which are theoretically less well behaved and have
larger uncertainties.

%
\begin{figure}
 \centering
   \includegraphics[height=8.4cm]{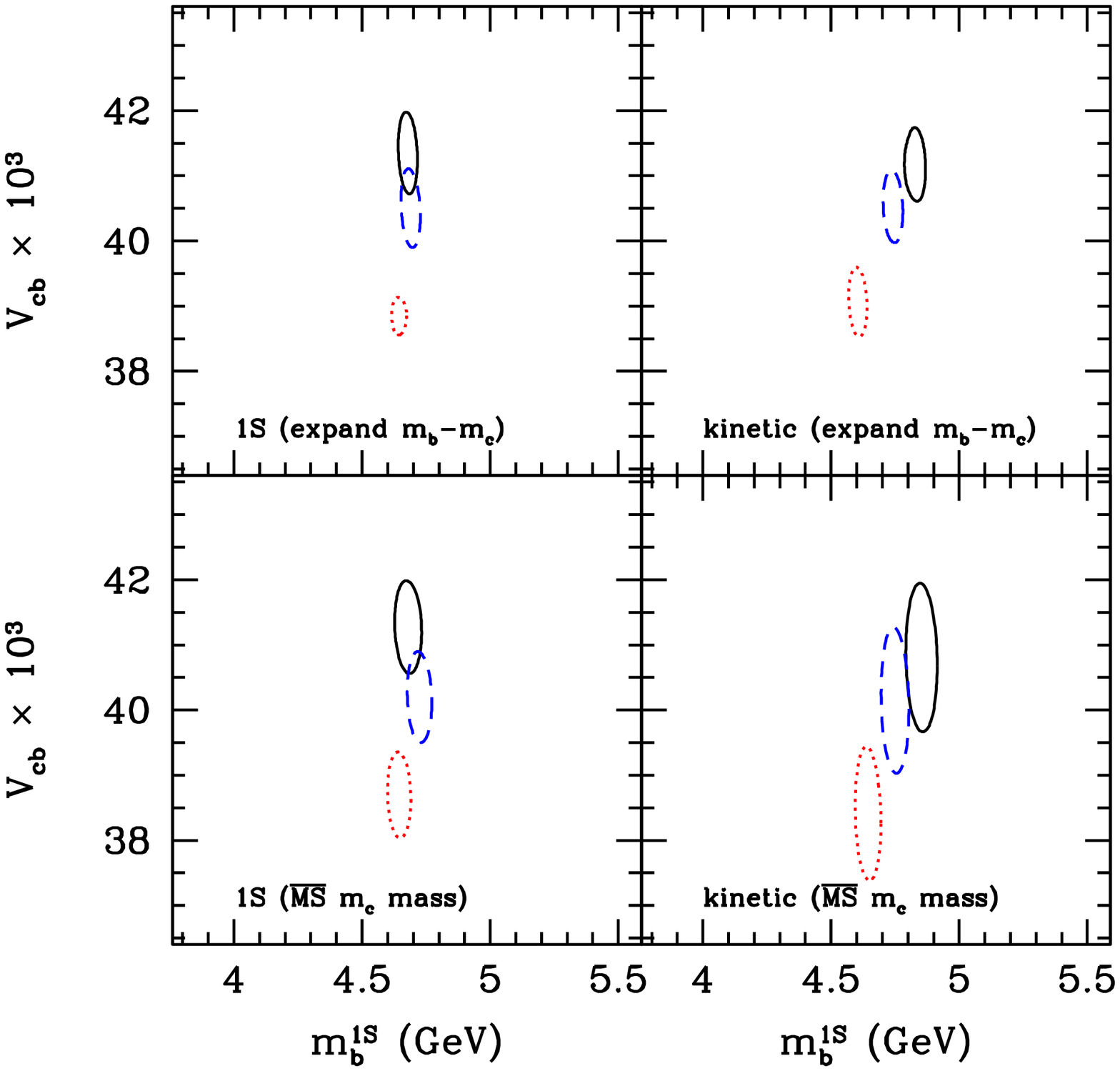}
\hspace{-0.35in}    \includegraphics[height=8.4cm]{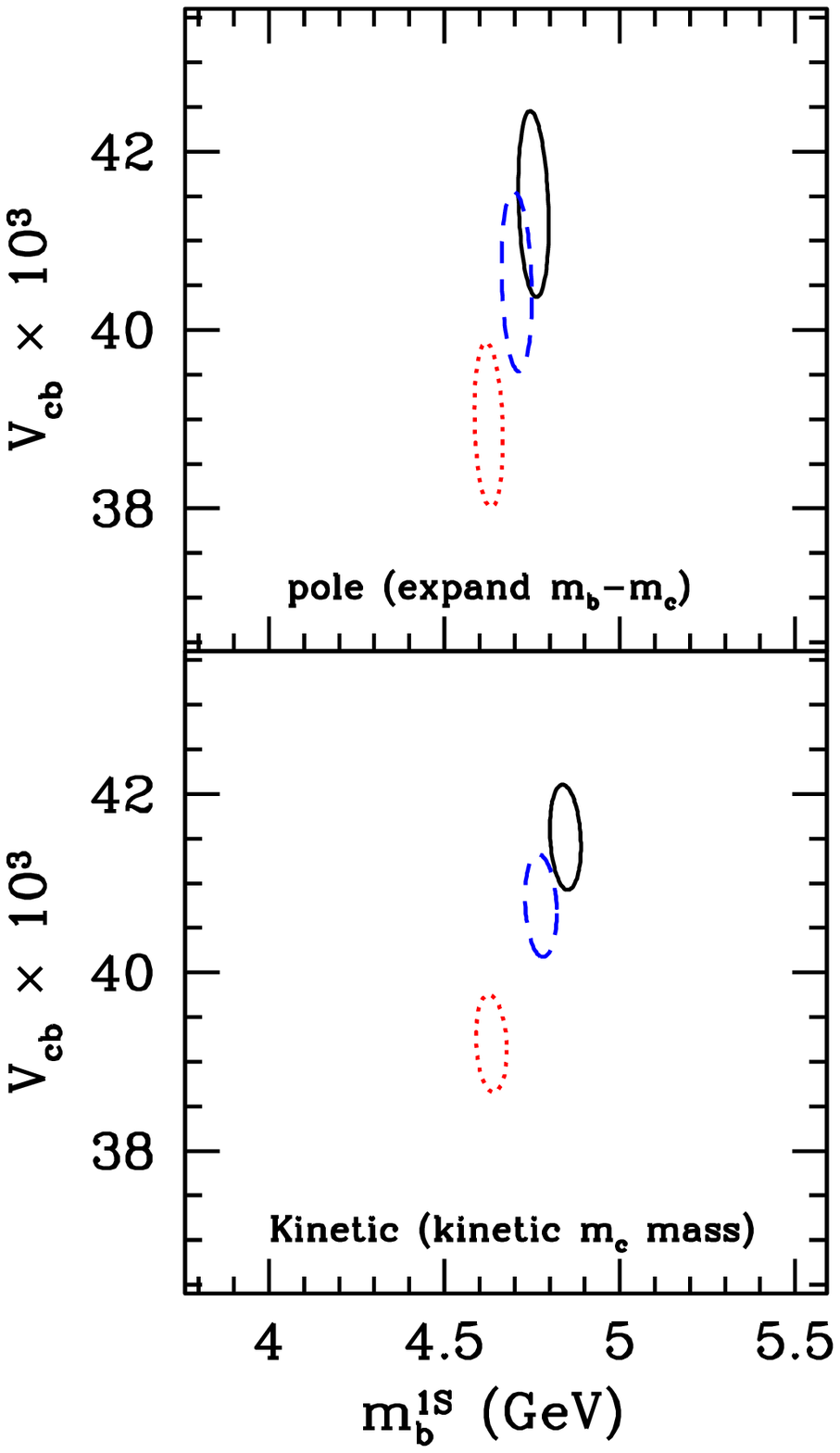}
 \caption{Convergence in $\alpha_s$ of the fit, in
different schemes. The dotted (red) dashed (blue) and solid (black)
ellipses denote the result of the fit when retaining terms of order
$\alpha_s^0$, $\alpha_s$ and $\beta_0\alpha_s^2$, respectively.}
\label{fig:bauer-alpha}
\end{figure}

The group of Bauer {\it et al} performs a global fit using data on
semileptonic and radiative decays from the BABAR, BELLE, CDF, CLEO, DELPHI
collaborations. They compare the two approaches, expanding or not in
$m_c$, and the different mass schemes. Fig.~\ref{fig:bauer-alpha}
shows the convergence in $\alpha_s$ of the computation in different
schemes. The dotted (red) dashed (blue) and solid (black) ellipses
denote the result of the fit when retaining terms of order
$\alpha_s^0$, $\alpha_s$ and $\beta_0\alpha_s^2$, respectively.

\begin{figure}
 \centering
    \includegraphics[height=8.4cm]{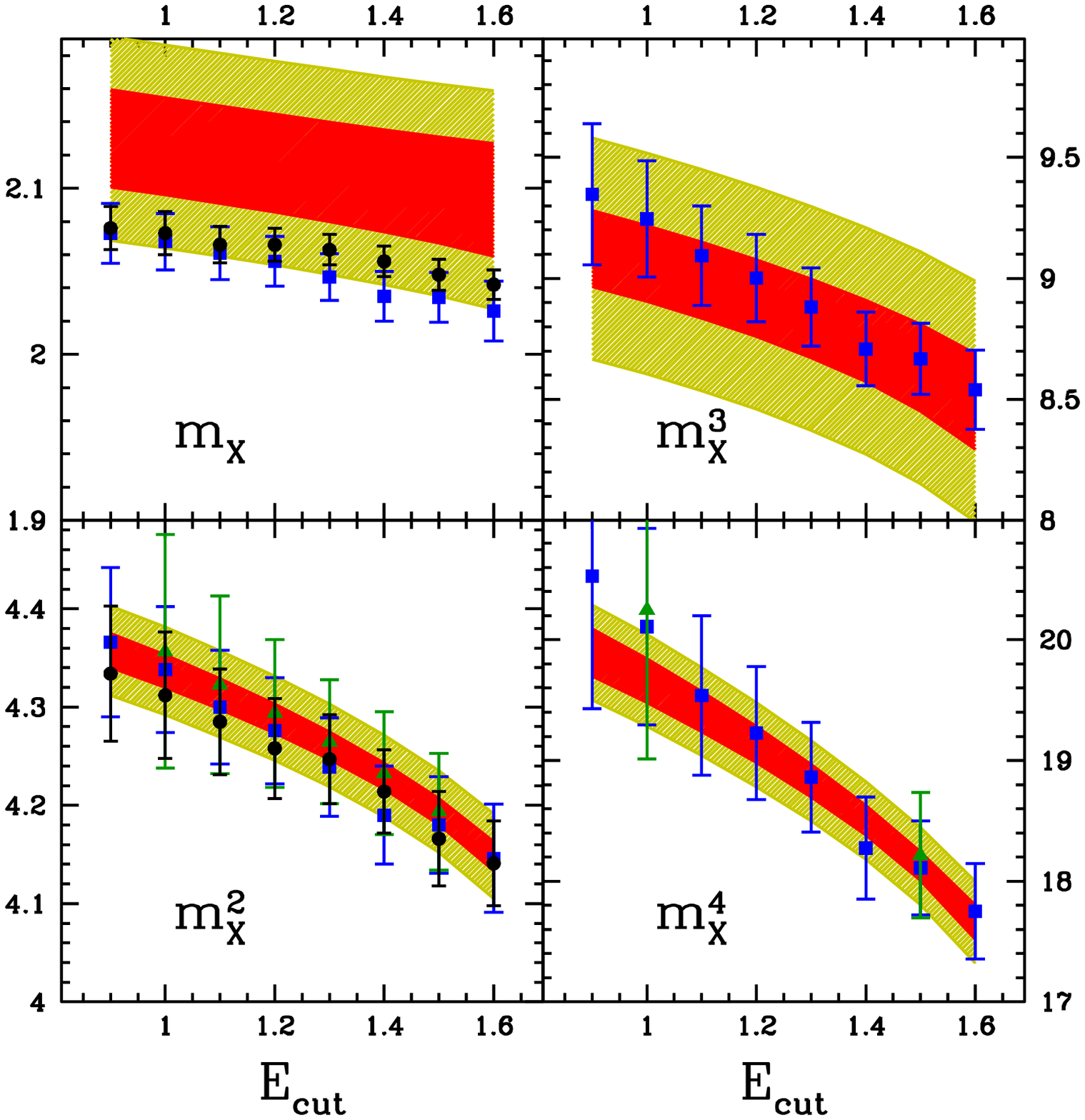}
    \includegraphics[height=8.4cm]{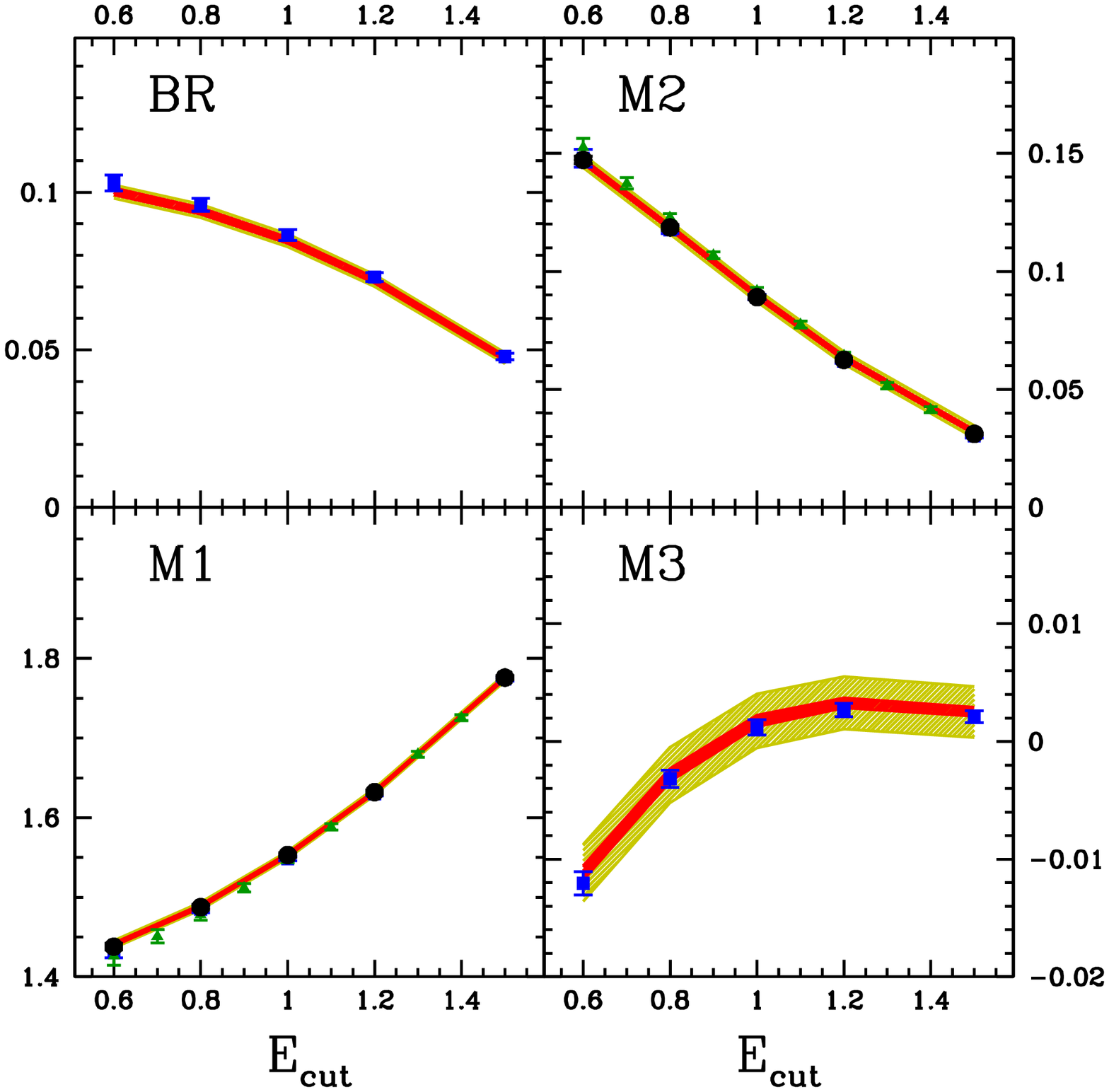}
\caption{Fit of moments as a function of $\ecut$. The data points  are from BABAR (squares), CLEO (triangles) and BELLE (circles).}
\label{fig:bauer-moments}
\end{figure}

%
\begin{figure}
 \centering
    \includegraphics[height=7.4cm]{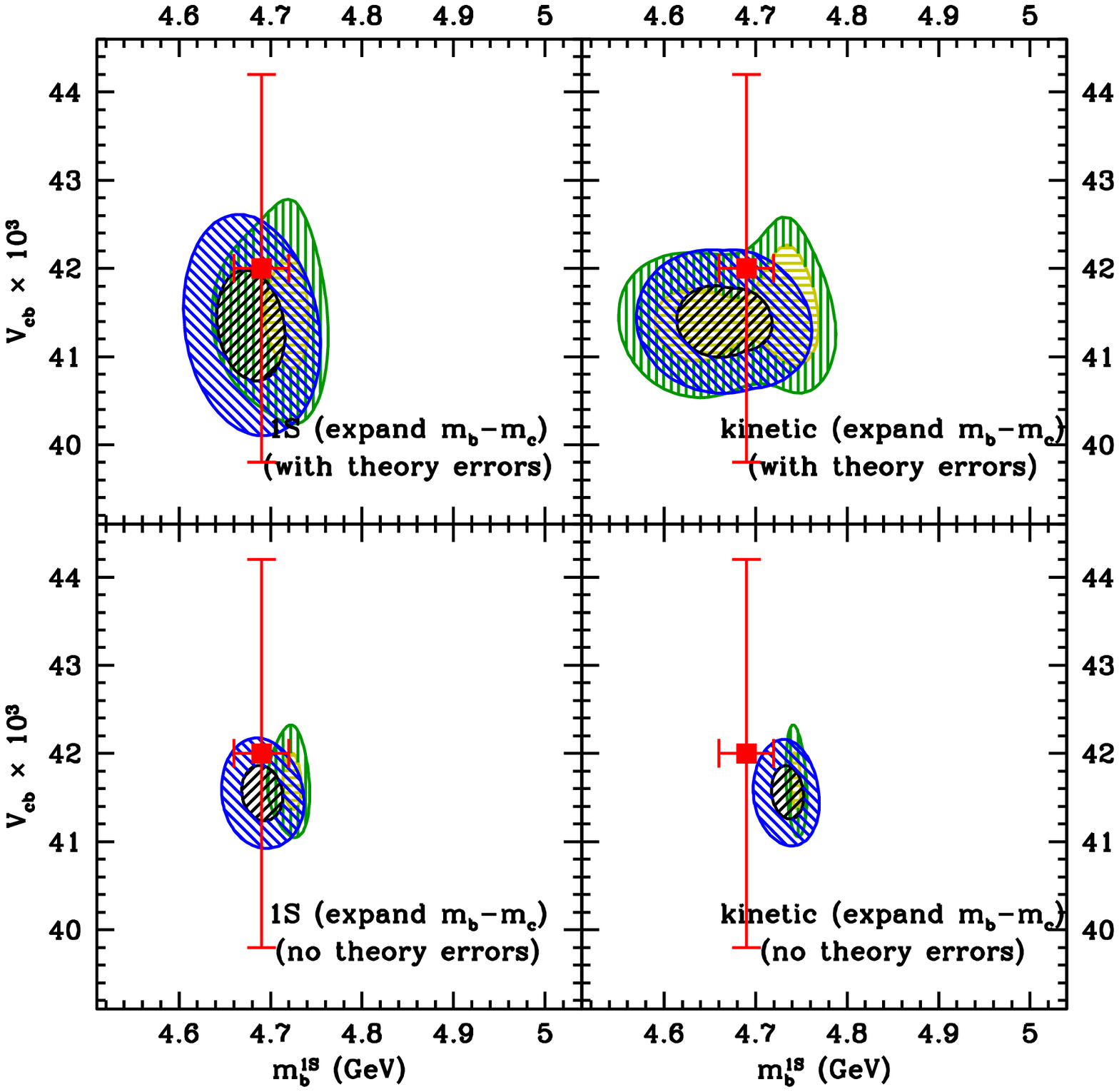}
\caption{Result of  fit for
$|V_{cb}|$ and $m_b$. The lower graphs do not include theory errors
and can be used to see the present theoretical limit of the extraction
of parameters. The ellipses denote regions of $\Delta\chi^2=1$ (black
and yellow), $\Delta\chi^2=4$ (blue and green). For the ellipses in
black and blue the parameters of order $(\LQCD)^3$ are restricted to
vary within $\pm(500~\text{MeV})^3$, while no such restriction is used
in the fit shown in yellow/green. The red cross shows the current PDG
value of $|V_{cb}|$ and the 1S $m_b$ determined from a fit to 
bottomonium\protect\cite{hoang}. }
\label{figbauerresults}
\end{figure}

Figure~\ref{fig:bauer-moments} shows the result of the fit, for
several moments as a function of $\ecut$. The data points are
superimposed: square (blue), triangles (green) and filled circles
(black) are from BABAR, CLEO and BELLE, respectively. The half integer
hadronic moments are shown, but not used in the fit for $\Vcb$. The
shaded region indicates the theory error, which has been estimated
from the size of the highest order retained in the double expansion in
$\alpha_s$ and $\LQCD/m_b$. The fit shown here is for the HQET/OPE
approach (with $1/m_c$ expansion) in the 1S mass scheme. The graphs
labeled BR and M1 show the branching fraction and the first lepton
energy moment, while M2 and M3 show the second and third lepton energy
central moments.  The remarkable agreement between theory and
experiment indicates that the expansion is working very well and, in
particular, that local duality works extremely well for these
decays. The result of the fit for $\Vcb$ and $m_b$ is shown in
Fig.~\ref{figbauerresults}. It is quite remarkable that the fitted
value of $m_b$ is comparable to that obtained from masses of
bottomonium and with comparable precision\cite{hoang}.

\section{$|V_{cb}|$ from Exclusives: $B\to D^*\ell\nu_\ell $ and $B\to
  D\ell\nu_\ell $}
The rates are given by the HQET-inspired parametrizations
\begin{eqnarray*}\label{rates}
  \frac{\d\Gamma(B\to D^*\ell\bar\nu)}{ \d w} &=& \frac{G_F^2 m_B^5}{ 48\pi^3}\, 
  r_*^3\, (1-r_*)^2\, \sqrt{w^2-1}\, (w+1)^2 \\
  && \times \left[ 1 + \frac{4w}{ 1+w} \frac{1-2wr_*+r_*^2}{ (1-r_*)^2} \right]
  |V_{cb}|^2\,{ \FDs{}^2(w)}  \\[5pt]
  \frac{\d\Gamma(B\to D \ell\bar\nu)}{ \d w} &=& \frac{G_F^2 m_B^5}{ 48\pi^3}\, 
  r^3\, (1+r)^2\, (w^2-1)^{3/2}\, |V_{cb}|^2\, { \FD^2(w) }
\end{eqnarray*}
in terms of $\FD$ and $\FDs$, which are simply combination of form
factors of $V-A$ charged currents. Here $w\equiv p_B\cdot
p_D^{(*)}/M_BM_D^{(*)}$, and $r^{(*)}=M_D^{(*)}/M_B$. At lowest order
in HQET ${ \FD(1)}={ \FDs(1)}=1$, and Luke's Theorem insures the
corrections are small, $\FDs(1)-1=\mathcal{O}(\LQCD/m_c)^2$.

In order to determine $|V_{cb}|$ we need an accurate theory of
$\FDs(1)$, a measurement of the rate as a function of $w$ and an
extrapolation of this measurement to $w=1$.  Analyticity and unitarity
tightly constrain the functional form of ${
\FDs(w)}$\cite{boyd}. Taylor expanding about $w=1$, \beq
\FDs(w)=\FDs(1)[1+\rho^2(w-1)+c(w-1)^2+\ldots] \eeqn the unitarity
constraint gives a relation between the slope, $\rho^2$, and
curvature, $c$\cite{Caprini:1995wq}.  Sum rules give lower bounds on these
parameters\cite{dorsten}, but the bounds are not experimentally
significant. It is interesting to note that at leading order in HQET
the slope and curvature parameters are equal for $\FD(w)$ and
$\FDs(w)$. Corrections have been estimated, $\rho_{\FD}^2 -
\rho_{\FDs}^2 \simeq0.19$, to be compared with the experimental fit,
$\rho_{\FD}^2 - \rho_{\FDs}^2 \simeq -0.22\pm0.20$\cite{rho2}. A more
precise determination of the slopes and curvatures would help (and may
tax) the
theoretical understanding of sub-leading form factors in HQET.

The main limitation in determining $\Vcb$ comes from
theoretical uncertainties in $\FDs(1)-1$. The situation has not
changed since the BaBar book was published, $\FDs(1)=0.91\pm
0.04$\cite{babarbook}. The current lattice determination has
comparable errors, $\FDs(1)=0.913\begin{smallmatrix}+0.024\\
-0.017\end{smallmatrix}\begin{smallmatrix}+0.017\\
-0.030\end{smallmatrix}$\cite{okamoto}. For a comparison of the
determinations of $\Vcb$ from inclusive and exclusive decays see
Ref.~\cite{elisabetta} in these proceedings.

\OMIT{
%
\begin{figure}
 \centering
    \includegraphics[width=6.5in]{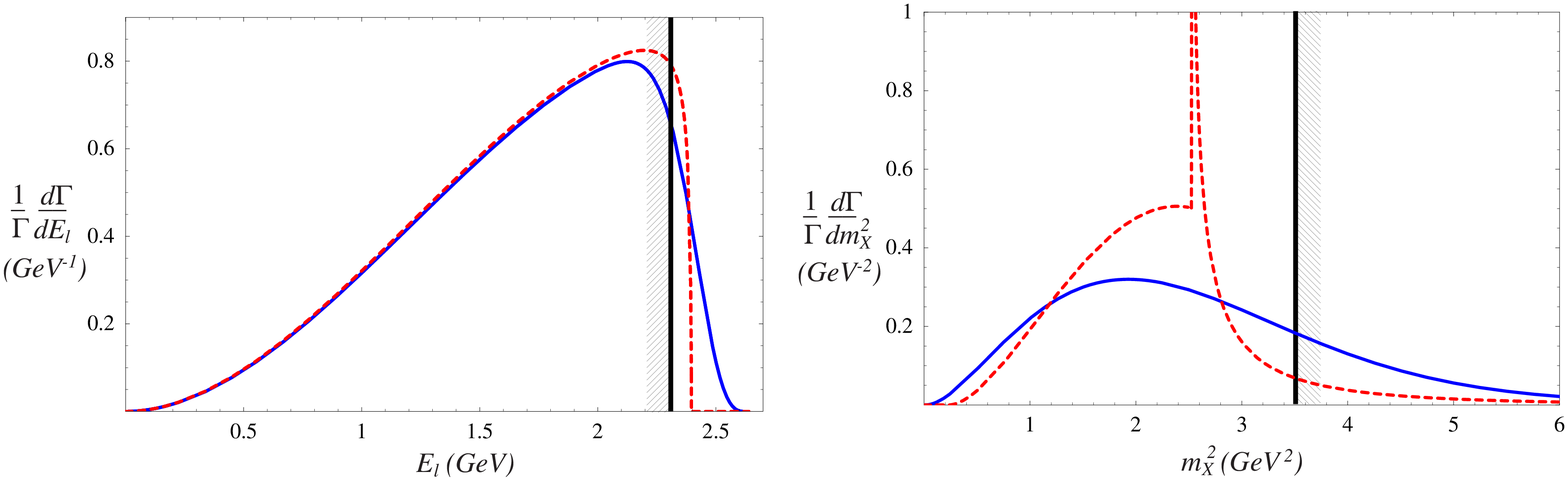}
\caption{Lepton energy and hadronic invariant mass spectra. The dashed curves are the $b$-quark decay results to $\CO(\alpha_s)$, while the solid curves are obtained by smearing with a model for the shape function\protect\cite{Bauer:2001gv}.  The vertical lines indicate the threshold of the region free from charm background.  }
\label{shapefunctioneffect}
\end{figure}
}

  \section{$|V_{ub}|$ from inclusive $B\to X_u\ell\nu_\ell$} The
theory of inclusive $b\to u\ell\nu$ processes is the same as in $b\to
c\ell \nu$. However, a straightforward application of theory is not
possible at the moment since experimentally the full spectrum is not
available. The problem is that the signal is swamped by $B\to
X_c\ell\nu_\ell$. The experimental solution: impose kinematic cuts to
suppress the $b\to c$ background. By imposing either, $m_X<m_D$,
$q^2>(m_B-m_D)^2$ or $E_\ell>(m_B^2-m_D^2)/2m_B $ one restricts the
measurement to events which are not kinematically accessible to $b\to
c$ decays. However, the experimental solution is largely incompatible
with theory\cite{Bauer:2001gv}. The endpoint spectra in $B\to X_u\ell\nu_\ell$ is given
in terms of a non-perturbative ``shape function,'' $f(x)$. One
approach is to measure $f(x)$ in $B\to X_s\gamma$ and use it to
eliminate uncertainties in the determination of $\Vub$ from the $B\to
X_u\ell\nu_\ell$ analysis\cite{neubert}. However, the universality of
the shape function is violated at order $1/m_b$, and the size of these
corrections is uncertain\cite{stewart}.

The sensitivity to the shape function is least with the cut on the
lepton invariant mass, $q^2>(m_B-m_D)^2$, but the cut picks up a small
region of the Dalitz plot. The hadronic mass cut includes a large
fraction of the Dalitz plot, but is more sensitive to $f(x)$. For now,
the best results are obtained by combining both cuts, trying to
maximize the rate while keeping dependence on $f(x)$ to a
minimum. Theoretical estimates of the uncertainty in the determination
of $\Vub$ from this method are in the 10\% range\cite{bauerluke}.

\section{$|V_{ub}|$ from Exclusive Decays}
\begin{figure}
\centering
  \includegraphics[height=6.0cm]{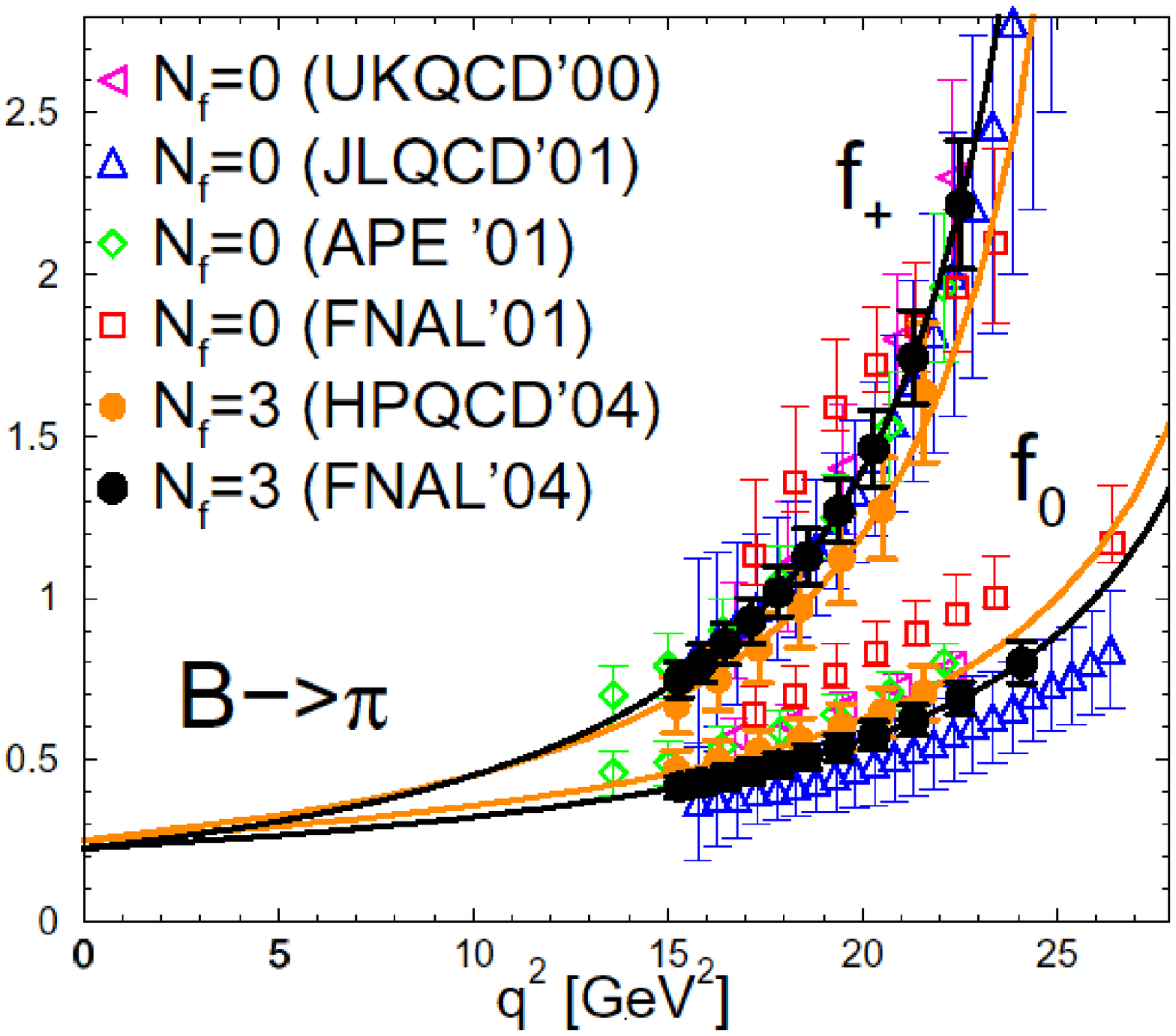}
\caption{Form factors for $B\to \pi\ell\nu$ computed by simulations of
  lattice QCD\cite{okamoto}. }
\label{fig:B->pi}
\end{figure}
The determination of $\Vub$ from exclusive decays requires {\it a
priori} knowledge of the form factors for $B\to\pi$ and/or $B\to\rho$,
{\it e.g.,}
\beq
\vev{\pi(p_\pi)|\bar b\gamma_\mu q| B(P_B)}
= \left((p_B+p_\pi)_\mu-\frac{m_B^2-m_\pi^2}{q^2}q_\mu\right){ F_+(q^2)}
+\frac{m_B^2-m_\pi^2}{q^2}q_\mu { F_0(q^2)}\,.
\eeqn
HQET does {\it not} fix normalization at zero recoil as it does in $B\to
D^{(*)}\ell\nu$.  Analyticity and unitarity do constrain the
functional form, which helps, {\it e.g.,}  to interpolate lattice results.
An interpolation of lattice results is shown in
Fig.~\ref{fig:B->pi}. One can use this to determine $\Vub$, but the
errors are large. Belle finds, from $B\to\pi\ell\nu$ restricted to
$q^2>16\text{GeV}^2$, $\Vub=(3.87\pm0.70\pm0.22 ^{+0.85}_{-0.51}) \times
10^{-3}$ and $\Vub= (4.73 \pm 0.85 \pm 0.27 ^{+0.74}_{-0.50}) \times 10^{-3}$
using lattice results from FNAL'04 and HPQCD,
respectively\cite{Abe:2004zm}.
The theory errors,  
$\sim \pm20$\%, are not expected to be significantly reduced soon. There has
also been some effort to understand the precision with which $\Vtd$
can be determined from $B\to\rho\gamma$ decays\cite{Grinstein:2000pc}.

New ideas are needed to reduce the error on $\Vub$ from exclusive
decays to the sub-10\% level, hopefuly to a few percent. One recent
proposal\cite{ligetiwise} is to use double ratios\cite{doubleratios}
to eliminate hadronic uncertainties. A double-ratio is a ratio of
ratios of quantities that are fixed by two symmetries. For example, 
\beq 
R_1={ \frac{f_{B_s}/f_{B}}{f_{D_s}/f_{D}}} = { \frac{f_{B_s}/f_{D_s}}{f_{B}/f_{D}}} = 1 
     + {\cal O}\left(m_s\left(\frac{1}{m_c}-\frac{1}{m_b}\right)\right)
\eeqn
The ratios $f_{B_s}/f_{B}$ and $f_{D_s}/f_{D}$ are unity in the
flavor-$SU(3)$ limit, while the ratios $f_{B_s}/f_{D_s}$ and
$f_{B}/f_{D}$ are fixed to $\sqrt{m_c/m_b}$ by heavy-quark
symmetry. Hence $R_1-1$ vanishes either as $m_s\to 0$ or as
$m_Q\to\infty$, as indicated.

To use a double-ratio in a precise determination of 
$|V_{ub}/V_{tb}V_{ts}|$, one measures, for $4m_c^2<q^2\lesssim
q^2_{\rm max}=(m_B-m_{\rho,K^*})^2 $
\beq
\frac{\mbox{d}\Gamma(\bar B\to \rho e\nu)/\mbox{d}q^2}
{\mbox{d}\Gamma(\bar B\to K^* \ell^+ \ell^-)/\mbox{d}q^2}
= 
{ \frac{|V_{ub}|^2}{|V_{tb} V_{ts}^*|^2}} \cdot \frac{8\pi^2}{\alpha^2}\cdot
\frac{1}{N_{\rm eff}(q^2)}
\frac{\sum_\lambda |H_\lambda^{B\to\rho}(q^2)|^2}
{\sum_\lambda |H_\lambda^{B\to K^*}(q^2)|^2}\,.
\eeqn
Here  $H_\lambda $ are helicity
amplitudes, and $N_{\rm eff}(q^2)$ is a calculable function that
incorporates the effects of penguin diagrams. In addition one must 
measure  decay spectra for $D\to \rho\ell\nu $ and $D\to K^*\ell\nu $
and express all rates as functions of $y=E_V/m_V$ ($V=\rho,
K^*$). Then one uses double-ratio magic. Let
\beq
R_{B\to V}(y) \equiv \frac{\sum_\lambda |H_\lambda^{B\to\rho}(y)|^2}
{\sum_\lambda |H_\lambda^{B\to K^*}(y)|^2} 
\qquad\qquad R_{D\to V}(y) \equiv  \frac{\sum_\lambda |H_\lambda^{D\to\rho}(y)|^2}{\sum_\lambda |H_\lambda^{D\to K^*}(y)|^2}
\eeqn
then 
\beq
R_{B\to V}(y) =R_{D\to V}(y)\Big(1 + {\cal O}(m_s(\frac{1}{m_c}-\frac{1}{m_b}))\Big) 
\eeqn
can be used to eliminate the unknown helicity amplitudes from the
analysis.

For the calculation of the factor $ N_{\rm eff}(q^2) = |C_9^{\rm eff}
+ \frac{2m_b^2}{q^2} C_7^{\rm eff}|^2 + |C_{10}|^2 + O(\Lambda/m_b)$,
one may include the effects of penguin diagrams through an OPE. This
is indicated above through the superscript ``eff'' ($C_{7,9}\to
C_{7,9}^{\rm eff}$). In applying the OPE,  there is no need to assume that the four quark operators factorize into the product of two currents. However, the OPE is used in the timelike region, so this assumes local duality. The situation
is under fairly good control, however, since the penguin contribution
is very small compared to the leading contribution from the contact
terms. Although the contribution is small, it is crucial to include it in order to minimize the scale dependence of  $N_{\rm eff}(q^2)$. The uncertainty in the scale dependence in $N_{\rm eff}(q^2)$,
which should be an RG-invariant, is only a couple of percent in a NNLO calculation.

Although a comprehensive study of the corrections has not been
performed, preliminary estimates indicate that the theoretcial error
in the determination of $\Vub$ could well be bellow 10\% with this
method. For example, the deviation of $R_1$ from 1 is about 2\% in
model and lattice calculations\cite{Cvetic:2004qg}. Similar results
are obtained for the double ratio of form factors for $B/D\to K^*/
\rho$\cite{Ligeti:1997aq}.  

\section*{Acknowledgments}
This work is supported in part by a
grant from the Department of Energy under Grant DE-FG03-97ER40546.

%
\label{GrinsteinEnd}

\end{document}